# Effect of internal resonance on the dynamics of MoS$_2$ resonator

Nishta Arora[1]* and A.K. Naik[1]*

[1]*Centre for Nanoscience and Engineering, Indian Institute of Science, Bangalore 560012, India*

E-mail: nishta@iisc.ac.in, anaik@iisc.ac.in

**Nonlinear modal interactions and associated internal resonance phenomena have recently been used to demonstrate improved oscillator performance and enhanced sensing capabilities. Here, we show tunable modal interaction in a MoS$_2$ resonator. We achieve the tunability of coupling between these initially uncoupled modes by using electrostatic gate voltages. This tunable coupling enables us to make the modes commensurate and observe energy exchange between the modes. We attribute the strong energy exchange between the vibrational modes to 2:1 internal resonance. This interaction strongly affects the dynamics of the modal response of such resonators. We observe peak splitting, a signature of energy exchange between the modes even when the modal response is in the linear regime. We model our device to explain the observed effect of excitation, detuning of modal frequencies, and intermodal coupling strength on the resonator dynamics. MoS$_2$ resonators explored in this work are ideal for understanding the rich dynamics offered through the intermodal coupling.**

Micro and nanoelectromechanical resonators are used for various applications including sensing[1,2,3], ultrastable oscillators for timing and frequency control[4,5], for fundamental studies in nonlinear dynamics[6,7] and quantum science[8,9,10,11]. The technological revolution towards miniaturization and advancement in fabrication techniques has led to the development of arrays of resonators[12], which are inherently coupled[13,14,15]. There is immense interest in broadening their applications and discovering new functionalities. Recently, coupled resonators have been shown to enhance the sensitivity of mass sensors[16,17] and hold promise for future nanomechanical technologies such as logic[18], memory[19], and computing[20,21]. However, to exploit a coupled network of resonators, it is essential to understand the influence of coupling on the dynamics of such resonators. Furthermore, the coupling can also be present between multiple modes of a single resonator[22,23,24]. The mode coupling mechanisms based on mechanical linkages, parametric pumping[22,25], dielectric coupling[26], and internal resonance[4,27] have been investigated. This coupling has been exploited in mechanical resonators to observe phenomena such as phonon lasing[28,29], cooling, amplification[30,31,32] and state squeezing[33,34]. However, our understanding of modal coupling through internal resonance is limited as it requires a meticulous design of device parameters to obtain resonant modes that are commensurate. Previous studies based on micro-resonators[35,36] offer weak frequency tunability and is a crucial challenge limiting its practical applications.

In contrast, two-dimensional materials such as graphene and metal dichalcogenides have highly tunable resonant frequencies[37,38]. This post-fabrication frequency tunability enables internal resonance condition to be easily satisfied during the measurements. Furthermore, these ultrathin devices are easy to scale-up for large scale production[39,40] using chemical vapor deposition (CVD) techniques, making them ideal for scientific and technological applications that exploit mode coupling. The coupling between modes of two-dimensional resonators[41,42] and carbon nanotubes[43,44] through internal resonance[45] has been demonstrated in strongly driven resonators exhibiting nonlinear vibrations. This coupling through internal resonance enables energy transfer between vibrational modes even if resonant frequencies are far apart in the frequency range. The necessary condition is that the ratio of resonant frequencies of the modes be commensurate[45]. In this work, we demonstrate strong intermodal coupling due to 2:1 internal resonance in $MoS_2$ drum resonators. We observe this signature of inter-modal coupling in the linear regime previously reported in micromechanical beams[46], but the same has not been observed before for ultrathin two-dimensional resonators. We demonstrate that the nonlinearity of the mode is not an essential condition to observe strong intermodal coupling through internal resonance. Since most sensing applications are performed in the linear regime, it is imperative to understand the effect of intermodal coupling in the linear regime. We can tune this coupling through electrostatic control provided by the gate electrode.

The simplified measurement circuit of two source mixed down ($1\omega$) technique used to actuate and transduce mechanical motion is shown in Figure1(a). In this technique[47,48], an RF signal ($V_g^{ac}$) with frequency $\omega$ and a DC bias ($V_g^{dc}$) is applied at the gate to actuate the resonator. Another RF signal ($V_s^{ac}$) with frequency $\omega + \delta\omega$ is applied at the source. Thus, the source-drain current and channel carrier density are modulated at frequency $\omega + \delta\omega$ and $\omega$, respectively. This modulation at different frequencies leads to mixed down current at $\delta\omega$, which is detected using a lock-in amplifier. At the resonance frequency, the displacement is maximum, resulting in a large change in the carrier density and the current. We carry out all the measurements at 370K and a vacuum below $10^{-6}$ Torr. The scanning electron micrograph (SEM) of a few layered MoS$_2$ resonator is shown in Figure 1(b). The detailed fabrication process is provided in the supplementary information (S1). Figure 1(c) shows the dispersion of the resonant frequency of the device with the applied DC gate bias ($V_g^{dc}$). The measurement was carried out at a source-drain bias ($V_s^{ac}$) of 20mV and AC gate bias ($V_g^{ac}$) of 50mV. The abrupt change in the resonant frequency (at $V_g^{dc} \approx -22V$ and $V_g^{dc} \approx 20V$) is attributed to clamping instability[49] (see supplementary section S2 for more discussion). Similar dynamical responses are reported in CNT and nanowire-based mechanical resonators[49,50]. All measurements reported here are for DC gate voltages ($V_g^{dc}$) beyond 20V, and the modes of the device are labeled as shown in Figure 1(c).

Using the electrostatic gate voltage ($V_g^{dc}$), we can tune the resonant frequency of the modes of the device. The tunability of the resonant frequency of each mode with the DC gate voltage is slightly different. This tunability enables us to modify the ratio of the resonant frequencies to achieve internal resonance conditions. Figure 2(a) depicts the frequency response curve of mode B (with resonance frequency $\omega_1$) and mode C (with resonance frequency $\omega_2$) at $V_g^{dc} = 21V$ with increasing AC drive voltages ($V_g^{ac}$). The response curves are vertically offset for different $V_g^{ac}$ drives (except 20mV) for visual clarity. The source-drain bias ($V_s^{ac}$) for all measurements is fixed at 20mV. The frequency response of mode B shows a normal linear Lorentzian response for small AC drives (20mV and 25mV). As we increase the $V_g^{ac}$, the response deviates from a simple Lorentzian, and the resonance peak splits into two. This splitting grows with the increasing $V_g^{ac}$. The corresponding phase response in the frequency range from 15 to 25 MHz (see supplementary S3) indicates that there are no other resonant modes nearby. The splitting in the response peak can be visualized as a dip appearing at $\omega_d \approx \omega_2/2$. The splitting is the result of mode coupling due to internal resonance. The corresponding exchange of energy between the two modes has been observed earlier in cantilevers and clamped-clamped beams[35,51,52]. But similar results have not been observed for two-dimensional materials in the linear regime. At $V_g^{dc} = 21$ V, the mode C (also shown in Figure2 (a)) has the resonance frequency at approximately twice the resonant frequency of mode B. Hence, we attribute the splitting to the internal resonance of 2:1. It is important to note that the resonator response is in the linear regime (see supplementary S4 for more details). The energy exchange between the modes is the result of the quadratic nonlinearity of the coupling term and not due to the nonlinearity of the mode itself[41]. The coupling arises due to membrane tension modification at high gate voltages leading to nonlinear modal interaction among multiple modes[29]. We have done similar measurements at $V_g^{dc} = 23V$ (Figure 2(b)) and $V_g^{dc} = 25V$ (Figure 2(c)). In both these cases, we do not observe any splitting in the peaks in either the amplitude or the phase response of the mode B (supplementary information S3). We attribute this to the different frequency tunings for the two modes with the applied DC gate voltages. As the gate bias is changed, the system is detuned away from the internal resonance condition ($\omega_2/\omega_1 \approx 2.00$). The resonance frequency ratio of the two modes for all three gate voltages (21V, 23V, and 25V) is shown in Figure 3. For DC gate voltages of 23V and 25V, the ratio of resonant frequencies is detuned away from the ratio of 2:1. Due to this detuning, we do not observe the intermodal energy exchange between mode B and C and the associated peak splitting in mode B.

To identify the effect of excitation, detuning of resonant frequencies ($\omega_2/\omega_1$) from internal resonance condition and the effect of coupling strength, we perform numerical simulations. We model our system as two linear resonators coupled through the nonlinear quadratic coupling. The quadratic coupling arises due to the interaction Hamiltonian, $H_{int} = x_1^2 x_2 + x_1 x_2^2$. We choose the Hamiltonian such that the interaction potential defining the energy exchange between the two modes is symmetric[29]. The equation of motion for our system (see supplementary S5 for more details) is:

$$\ddot{x}_1 + \omega_1^2 x_1 + 2\mu_1 \dot{x}_1 + \alpha_1 x_1^3 + \gamma_{22} x_2^2 + 2\gamma_{12} x_1 x_2 = \frac{F_1}{m_1} Cos(\omega_d t) \qquad (1)$$

$$\ddot{x}_2 + \omega_2^2 x_2 + 2\mu_2 \dot{x}_2 + \alpha_2 x_2^3 + \gamma_{11} x_1^2 + 2\gamma_{21} x_1 x_2 = \frac{F_2}{m_2} Cos(\omega_d t) \qquad (2)$$

where, $x_1, x_2$ are displacements of the first and second mode respectively, $\mu_1, \mu_2 \approx \mu$ is the modal damping coefficient, $\alpha_1, \alpha_2$ represents the effective Duffing nonlinearity of the two modes. The effective Duffing nonlinearity is the combination of tension produced in the membrane at large vibration amplitudes and the broken symmetry of the resonator caused due to the electrostatic force between the membrane and the gate. $\gamma_{12} \approx \gamma_{11}, \gamma_{21} \approx \gamma_{22}$ are the coupling coefficients, $\omega_1$ and $\omega_2$ are the resonant frequency of the two modes, respectively, $\omega_d$ is the excitation signal frequency at which force $F_1$ and $F_2$ are applied to the first and the second mode, respectively. We numerically evaluate the steady-state response of the two interacting modes under the given driving conditions. The parameters used and other simulation details are provided in the supplementary section S6.

Since the device operates in the linear regime, the nonlinear coefficients of the two modes $\alpha_1, \alpha_2$ do not have any significant role in the dynamics of the resonator. Hence, in all further simulations to reduce the complexity, we assume $\alpha_1, \alpha_2 = 0$. Also, in the experimental data presented here, a single mode (either mode B or mode C) is driven at a time. We simulate the same using the above equations by considering $F_2$ to be zero and vice-versa. In the absence of any coupling ($\gamma_{11} \approx \gamma_{22} \approx \gamma_{12} \approx \gamma_{21} = 0$), driving one mode would lead to no transfer of energy from the driven to the undriven mode. The simulated response of the same using the above equation for both the modes is shown in the supplementary (section S6.1).

The simulated frequency response of the two coupled modes with frequencies $\omega_1 = 1.00$ and $\omega_2 = 2.05$ under increasing drive force, is shown in Figure 4(a). The frequency response curve is similar to the one obtained experimentally for mode B. In this case, coupling strength denoted by dimensionless parameters $\gamma_{12} \approx \gamma_{11} = 0.01$ and $\gamma_{21} \approx \gamma_{22} = 0.03$ is kept constant. The frequency response of the first mode is a single resonance peak at low driving forces, indicating minimal interaction with the second mode. Increasing the excitation force leads to increased energy exchange between the first mode and the second mode. This energy exchange is observed experimentally as splitting in the peak, which increases with excitation. The simulated frequency response of the second mode while driving the first mode is shown in Figure 4(b). It shows that as we increase the driving force of the first mode, the amplitude response of the coupled second mode also increases, thus confirming the energy transfer.

We also simulate the frequency response of the modes while forcing the second mode instead of the first. Since the energy transfer from the second mode to the first is not significant, the driven second mode does not show any peak splitting (Figure 4(c)). We attribute the absence of peak splitting while forcing the second mode to the inefficient energy transfer. Figure 4(d) shows the response of the weakly coupled first mode on driving the second mode. The minimal change in amplitude demonstrates inefficient energy transfer between the modes.

To understand the effect of detuned frequencies on the splitting and coupling with higher mode, we simulate the response of the first mode by varying $\omega_2$ (Figure 5(a)) while keeping other parameters ($\gamma_{12} \approx \gamma_{11} = 0.01$ and $\gamma_{21} \approx \gamma_{22} = 0.03$) and force ($F_1 = 2.00$) constant. For the case of $\omega_2/\omega_1 = 2.00$, we observe symmetric splitting in the frequency response (blue curve), whereas when the ratio is detuned such that $\omega_2/\omega_1 < 2.00$ (orange curve), the splitting becomes asymmetric. The asymmetry in peak splitting can be reversed to the other direction by detuning the resonant frequency such that $\omega_2/\omega_1 > 2.00$. This asymmetric splitting is depicted in Figure 5(a) by the yellow curve. The corresponding frequency response of the second mode for the three detuning cases is shown in supplementary (section S6.2). The observed peak splitting is reduced by detuning the resonance frequencies away from the internal resonance condition ($\omega_2/\omega_1 = 2.00$). Beyond a certain detuning, no peak splitting is observed. This implies that the energy exchange between the two modes is not efficient enough to cause peak splitting through internal resonance.

Figure 5(b) illustrates the effect of coupling strength on the frequency response of mode 1. The peak splitting, a signature of intermodal coupling and energy transfer between coupled modes, becomes dominant at higher coupling strength. We do not observe any peak splitting for lower coupling strengths. This indicates that the energy exchange between the two modes is not strong enough to result in a peak splitting. Thus, to observe the strong effect of mode coupling on the dynamics of the resonator, it is imperative to have high modal coupling strength and commensurate modes. A similar effect of coupling strength on the modal response of the first mode for the case of the further detuned ratio of resonant frequencies is shown in supplementary (section S6.3).

In conclusion, we have demonstrated strong intermodal coupling through 2:1 internal resonance in MoS$_2$ resonators. We observe the signature of energy transfer between the modes in the linear regime. We can tune the coupling between mechanical modes using the electrostatic gate voltages. The theoretical model presented in this work is in excellent agreement with the experimental measurements and qualitatively captures the experimentally observed dynamics in MoS$_2$ resonators. The model discussed here is not just limited to two-dimensional materials but can be applied to other coupled systems demonstrating internal resonance. It provides insights into the effect of forcing, detuning from internal resonance condition, and strength of intermodal coupling on the dynamics of such resonators. The observation of peak splitting in resonances can be used to predict the presence of higher-order modes that are otherwise difficult to measure experimentally.

**Acknowledgments**

We acknowledge financial support from Science and Engineering Research Board, DST India, through grant EMR/2016/006479 and DST Nanomission, India, through grant SR/NM/NS-1157/2015(G). N.A. acknowledges fellowship support under Visvesvaraya Ph.D. Scheme, Ministry of Electronics and Information Technology (MeitY), India. We are grateful for funding from MHRD, MeitY, and DST Nano Mission for supporting the facilities at the Centre for Nano Science and Engineering. We also acknowledge the usage of the Micro and Nano Characterization Facility and National Nano Fabrication Facility at the Centre for Nano Science and Engineering. We thank Dr. Chandan Samanta and Prof. A.M. Van der Zande for helpful discussions.

# References

[1] J. Moser, J. Güttinger, A. Eichler, M.J. Esplandiu, D.E. Liu, M.I. Dykman, and A. Bachtold, Nat. Nanotechnol. **8**, 493 (2013).

[2] M.S. Hanay, S. Kelber, A.K. Naik, D. Chi, S. Hentz, E.C. Bullard, E. Colinet, L. Duraffourg, and M.L. Roukes, Nat. Nanotechnol. **7**, 602 (2012).

[3] H.J. Mamin, D. Rugar, B.W. Chui, and R. Budakian, Nature **430**, 329 (2004).

[4] D. Antonio, D.H. Zanette, and D. López, Nat. Commun. **3**, 806 (2012).

[5] C.T.C. Nguyen, IEEE Trans. Ultrason. Ferroelectr. Freq. Control **54**, 251 (2007).

[6] I. Kozinsky, H.W.C. Postma, O. Kogan, A. Husain, and M.L. Roukes, Phys. Rev. Lett. **99**, 207201 (2007).

[7] R.B. Karabalin, M.C. Cross, and M.L. Roukes, Phys. Rev. B **79**, 165309 (2009).

[8] R. Riedinger, A. Wallucks, I. Marinković, C. Löschnauer, M. Aspelmeyer, S. Hong, and S. Gröblacher, Nature **556**, 473 (2018).

[9] C.F. Ockeloen-Korppi, E. Damskägg, J.M. Pirkkalainen, M. Asjad, A.A. Clerk, F. Massel, M.J. Woolley, and M.A. Sillanpää, Nature **556**, 478 (2018).

[10] C.B. Møller, R.A. Thomas, G. Vasilakis, E. Zeuthen, Y. Tsaturyan, M. Balabas, K. Jensen, A. Schliesser, K. Hammerer, and E.S. Polzik, Nature **547**, 191 (2017).

[11] E.E. Wollman, C.U. Lei, A.J. Weinstein, J. Suh, A. Kronwald, F. Marquardt, A.A. Clerk, and K.C. Schwab, Science **349**, 952 (2015).

[12] S. Dominguez-Medina, S. Fostner, M. Defoort, M. Sansa, A.-K. Stark, M.A. Halim, E. Vernhes, M. Gely, G. Jourdan, and T. Alava, Science **362**, 918 (2018).

[13] M.H. Matheny, J. Emenheiser, W. Fon, A. Chapman, A. Salova, M. Rohden, J. Li, M.H. de Badyn, M. Pósfai, L. Duenas-Osorio, M. Mesbahi, J.P. Crutchfield, M.C. Cross, R.M. D'Souza, and M.L. Roukes, Science **363**, eaav7932 (2019).

[14] S.B. Shim, M. Imboden, and P. Mohanty, Science **316**, 95 (2007).

[15] M.H. Matheny, L.G. Villanueva, R.B. Karabalin, J.E. Sader, and M.L. Roukes, Nano Lett. **13**, 1622 (2013).

[16] H. Zhang, J. Huang, W. Yuan, and H. Chang, J. Microelectromechanical Syst. **25**, 937 (2016).

[17] C. Zhao, G.S. Wood, J. Xie, H. Chang, S.H. Pu, and M. Kraft, Sensors Actuators, A Phys. **232**, 151 (2015).

[18] S.C. Masmanidis, R.B. Karabalin, I. De Vlaminck, G. Borghs, M.R. Freeman, and M.L. Roukes, Science **317**, 780 (2007).

[19] A. Yao and T. Hikihara, Appl. Phys. Lett. **105**, 123104 (2014).

[20] M.A.A. Hafiz, L. Kosuru, and M.I. Younis, J. Appl. Phys. **120**, 074501 (2016).

[21] M. Freeman and W. Hiebert, Nat. Nanotechnol. **3**, 251 (2008).

[22] P. Prasad, N. Arora, and A.K. Naik, Nano Lett. **19**, 5862 (2019).

[23] H.J.R. Westra, M. Poot, H.S.J. Van Der Zant, and W.J. Venstra, Phys. Rev. Lett. **105**, 117205 (2010).

[24] M.J. Seitner, M. Abdi, A. Ridolfo, M.J. Hartmann, and E.M. Weig, Phys. Rev. Lett. **118**, 254301 (2017).

[25] C.H. Liu, I.S. Kim, and L.J. Lauhon, Nano Lett. **15**, 6727 (2015).

[26] T. Faust, J. Rieger, M.J. Seitner, J.P. Kotthaus, and E.M. Weig, Nat. Phys. **9**, 485 (2013).

[27] C. Chen, D.H. Zanette, D.A. Czaplewski, S. Shaw, and D. López, Nat. Commun. **8**, 15523 (2017).

[28] I. Mahboob, K. Nishiguchi, A. Fujiwara, and H. Yamaguchi, Phys. Rev. Lett. **110**, 127202 (2013).

[29] R. De Alba, F. Massel, I.R. Storch, T.S. Abhilash, A. Hui, P.L. McEuen, H.G. Craighead, and J.M. Parpia, Nat. Nanotechnol. **11**, 741 (2016).

[30] I. Mahboob, K. Nishiguchi, H. Okamoto, and H. Yamaguchi, Nat. Phys. **8**, 387 (2012).

[31] F. Sun, X. Dong, J. Zou, M.I. Dykman, and H.B. Chan, Nat. Commun. **7**, 12694 (2016).

[32] J.P. Mathew, R.N. Patel, A. Borah, R. Vijay, and M.M. Deshmukh, Nat. Nanotechnol. **11**, 747 (2016).

[33] Y.S. Patil, S. Chakram, L. Chang, and M. Vengalattore, Phys. Rev. Lett. **115**, 0172020 (2015).

[34] I. Mahboob, H. Okamoto, K. Onomitsu, and H. Yamaguchi, Phys. Rev. Lett. **113**, 167203 (2014).

[35] A.Z. Hajjaj, F.K. Alfosail, and M.I. Younis, Int. J. Non. Linear. Mech. **107**, 64 (2018).

[36] A.Z. Hajjaj, N. Jaber, M.A.A. Hafiz, S. Ilyas, and M.I. Younis, Phys. Lett. Sect. A Gen. At. Solid State Phys. **382**, 3393 (2018).

[37] P. Prasad, N. Arora, and A.K. Naik, Nanoscale **9**, 18299 (2017).

[38] M.M. Parmar, P.R.Y. Gangavarapu, and A.K. Naik, Appl. Phys. Lett. **107**, 113108 (2015).

[39] A.M. Van Der Zande, R.A. Barton, J.S. Alden, C.S. Ruiz-Vargas, W.S. Whitney, P.H.Q. Pham, J. Park, J.M. Parpia, H.G. Craighead, and P.L. McEuen, Nano Lett. **10**, 4869 (2010).

[40] C. Samanta, N. Arora, K.K. V., S. Raghavan, and A.K. Naik, Nanoscale **11**, 8394 (2019).

[41] C. Samanta, P.R. Yasasvi Gangavarapu, and A.K. Naik, Appl. Phys. Lett. **107**, 173110 (2015).

[42] S.S.P. Nathamgari, S. Dong, L. Medina, N. Moldovan, D. Rosenmann, R. Divan, D. Lopez,

L.J. Lauhon, and H.D. Espinosa, Nano Lett. **19**, 4052 (2019).

[43] A. Eichler, M. Del Álamo Ruiz, J.A. Plaza, and A. Bachtold, Phys. Rev. Lett. **109**, 025503 (2012).

[44] A. Castellanos-Gomez, H.B. Meerwaldt, W.J. Venstra, H.S.J. Van Der Zant, and G.A. Steele, Phys. Rev. B **86**, 041402 (2012).

[45] A.H. Nayfeh and D.T. Mook, *Nonlinear Oscillations* (Wiley-VCH, 1995).

[46] T. Zhang, C. Guo, Z. Jiang, and X. Wei, J. Appl. Phys. **126**, 164506 (2019).

[47] C. Chen, S. Rosenblatt, K.I. Bolotin, W. Kalb, P. Kim, I. Kymissis, H.L. Stormer, T.F. Heinz, and J. Hone, Nat. Nanotechnol. **4**, 861 (2009).

[48] I. Bargatin, E.B. Myers, J. Arlett, B. Gudlewski, and M.L. Roukes, Appl. Phys. Lett. **86**, 133109 (2005).

[49] M. Aykol, B. Hou, R. Dhall, S.W. Chang, W. Branham, J. Qiu, and S.B. Cronin, Nano Lett. **14**, 2426 (2014).

[50] H.S. Solanki, Semiconducting Nanowire Electromechanics, Tata Institute of Fundamental Research (Doctoral Thesis), 2011.

[51] A. Sarrafan, S. Azimi, F. Golnaraghi, and B. Bahreyni, Sci. Rep. **9**, 8648 (2019).

[52] T. Zhang, J. Ren, X. Wei, Z. Jiang, and R. Huan, Appl. Phys. Lett. **109**, 224102 (2016).

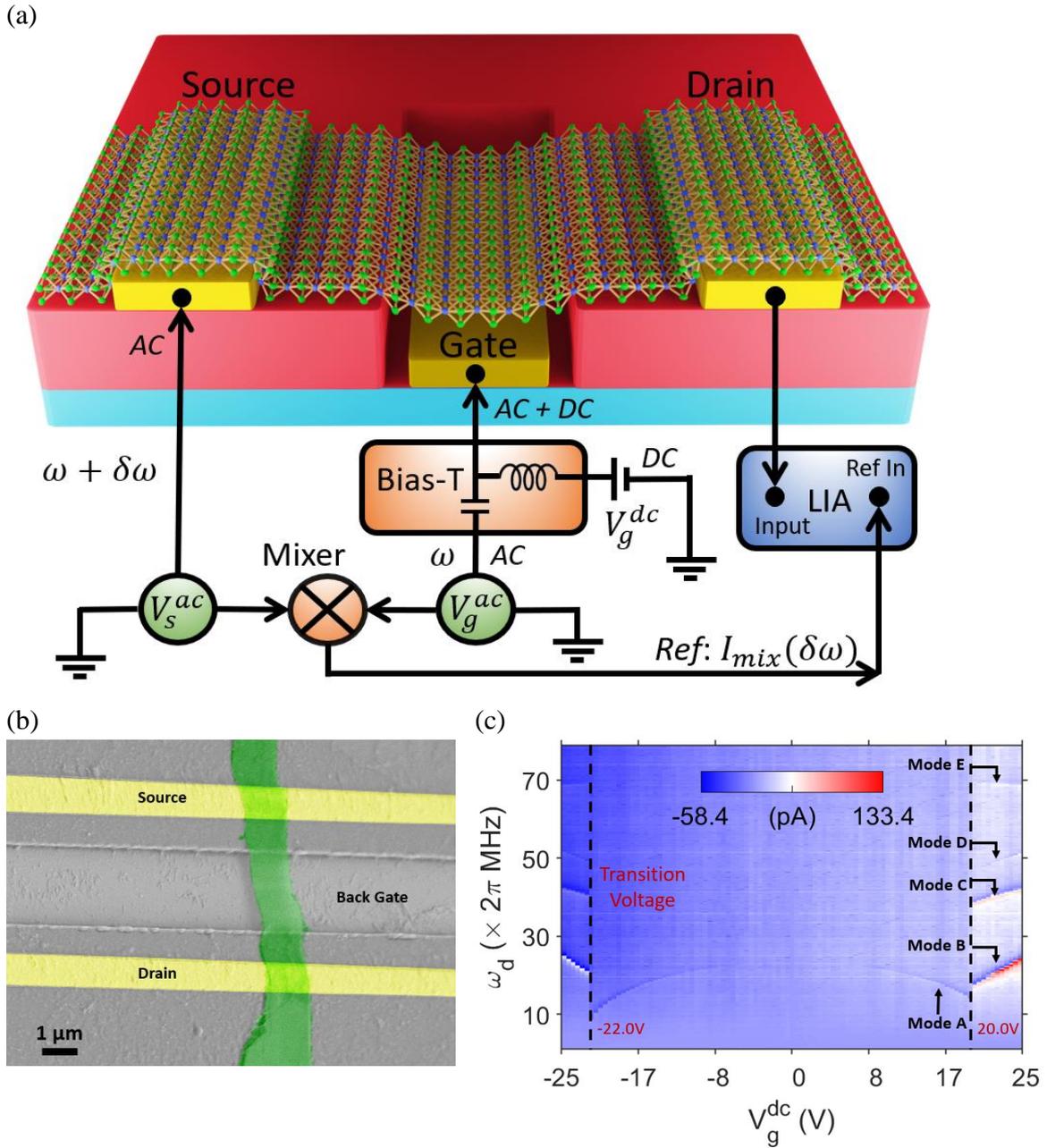

**Figure 1. (a)** Simplified schematic circuit of the experimental setup for actuation and detection using mixed down technique. **(b)** False colored scanning electron micrograph of MoS$_2$ resonator. The yellow lines represent the source and drain. The trench underneath the MoS$_2$ (green color) membrane with the gate electrode (back gate) is 300nm in depth. **(c)** Resonant frequency dispersion plot showing multiple resonance modes. The resonance frequencies are tuned through applied DC gate bias ($V_g^{dc}$).

(a)

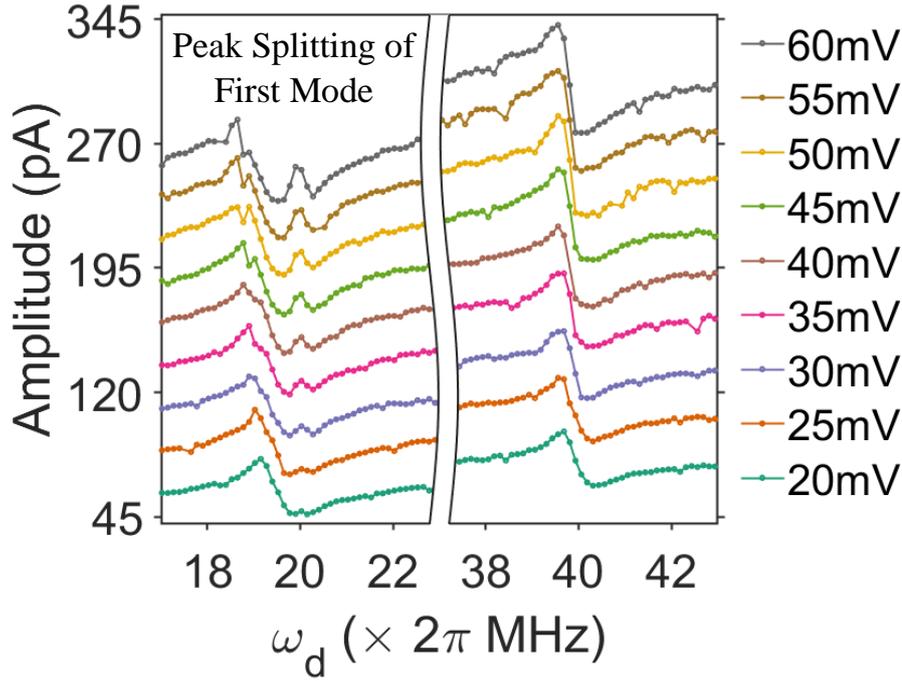

(b)

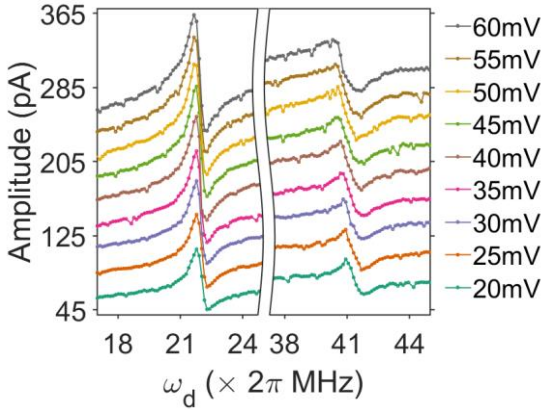

(c)

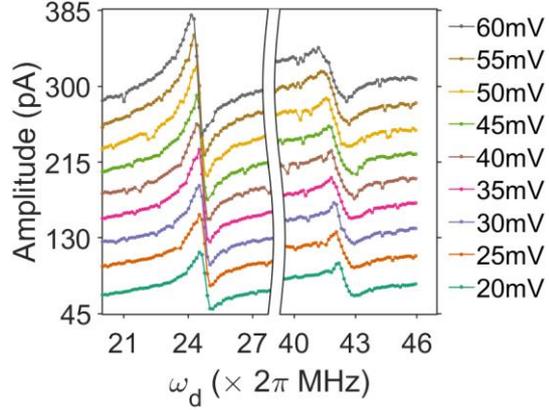

**Figure 2. (a)** Amplitude response of mode B and C observed at $V_g^{dc} = 21V$ gate bias with varying ac drive voltages ($V_g^{ac}$). Mode B starts to split at higher drive voltages due to modal coupling through internal resonance. **(b)** The amplitude response of mode B and C observed at $V_g^{dc} = 23V$ and **(c)** $V_g^{dc} = 25V$ gate bias with varying ac drive voltages ($V_g^{ac}$). No peak splitting is observed, since $\omega_2/\omega_1 \approx 2.00$ (2:1 internal resonance condition) is detuned by changing the gate bias ($V_g^{dc}$). The measurements were performed using ($1\omega$) technique with $V_s^{ac} = 20mV$ and mixdown frequency ($\delta\omega$) as $1987Hz$. The curves are offset for different $V_g^{ac}$ (except 20mV) for visual clarity.

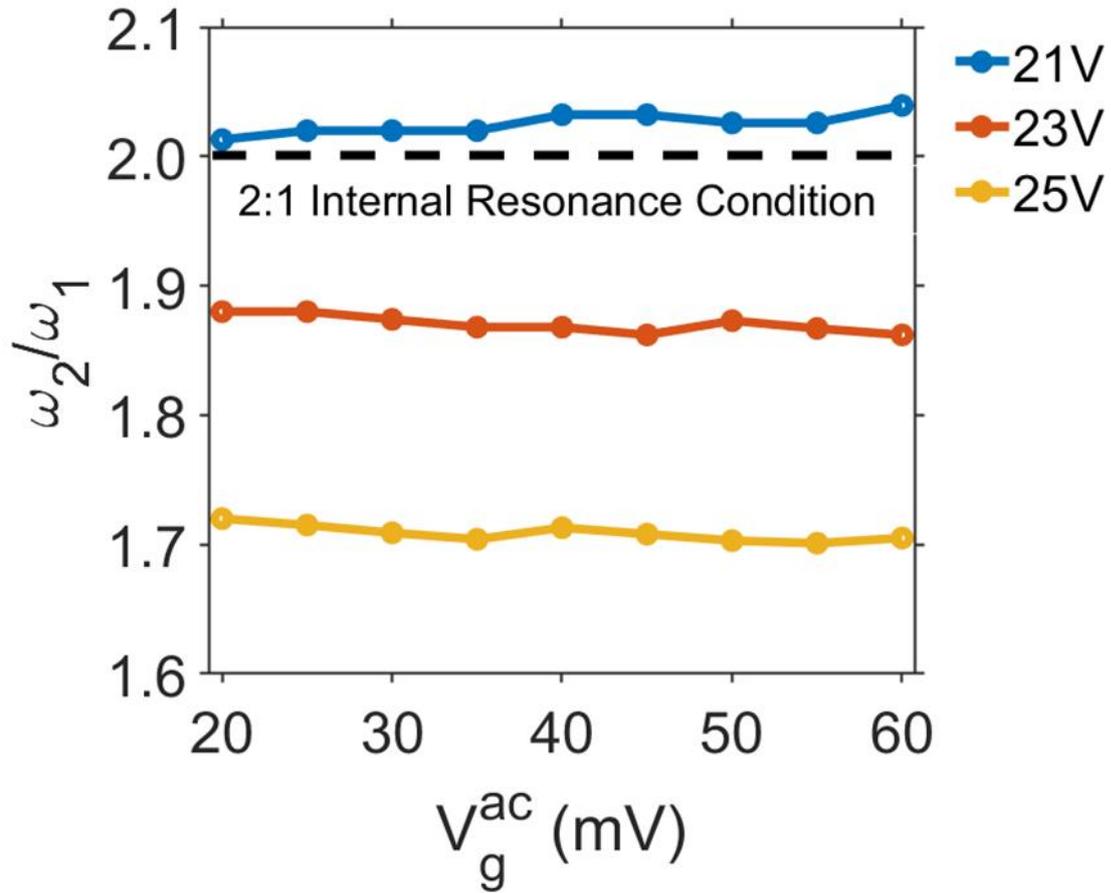

**Figure 3.** The ratio of resonant frequency of Mode C ($\omega_2$) and Mode B ($\omega_1$) with varying AC drives for three different DC gate voltages: $V_g^{dc} = 21V$ (blue), $V_g^{dc} = 23V$ (orange), $V_g^{dc} = 25V$ (yellow).

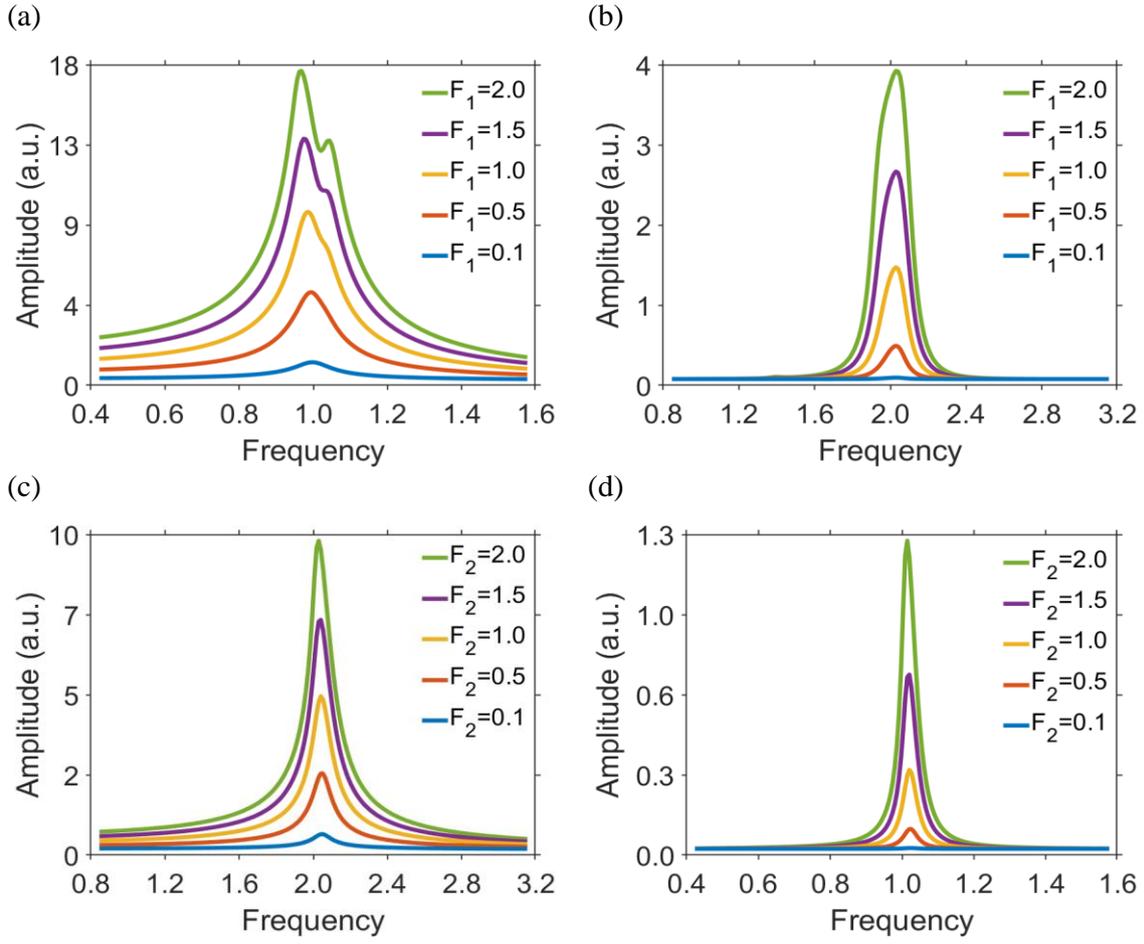

**Figure 4. Simulations:** Effect of increasing the driving force on the frequency response of **(a)** the first mode and **(b)** the second mode while forcing the first mode. At higher driving forces, peak splitting in the first mode is observed. Effect of increasing the driving force on the frequency response of **(c)** the second mode and **(d)** the first mode while forcing the second mode. Simulation parameters are kept constant throughout and are as follows: $\omega_1 = 1.00$, $\omega_2 = 2.05$, $\mu_1 \approx \mu_2 = 0.05$, $\gamma_{12} \approx \gamma_{11} = 0.01$, $\gamma_{21} \approx \gamma_{22} = 0.03$).

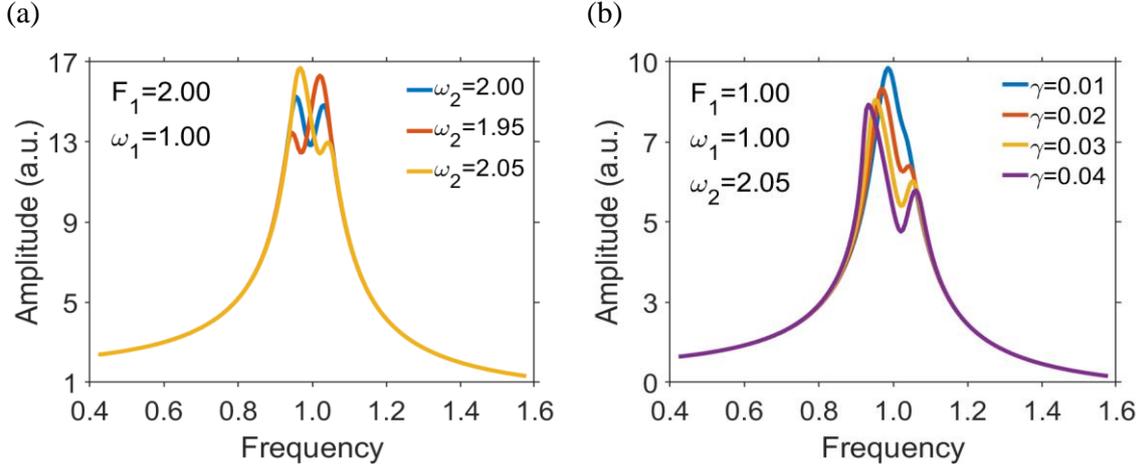

**Figure 5. Simulations: (a)** Effect of detuning of resonant frequencies. The ratio $\omega_2/\omega_1 = 2.00, 1.95, 2.05$ is changed by varying frequency of $\omega_2$ while keeping the force ($F_1 = 2.00$) and other parameters ($\gamma_{12} \approx \gamma_{11} = 0.01$ and $\gamma_{21} \approx \gamma_{22} = 0.03$) constant. The asymmetry in the splitting of the peak depends on the ratio of detuning from the internal resonance condition. **(b)** Effect of coupling strength on the frequency response of mode 1 ($\omega_1 = 1.00$) for mode coupled system with $\omega_1 = 1.00$ and $\omega_2 = 2.05$. Splitting in the peak is observed to increase with increasing coupling strength for the same forcing ($F_1 = 1.00$). For simplicity, we assume coupling as ($\gamma_{12} \approx \gamma_{11} \approx \gamma_{22} \approx \gamma_{21} = \gamma$).

Supporting Information

# Effect of internal resonance on the dynamics of MoS$_2$ Resonator

Nishta Arora[1] and A.K. Naik[1]

[1]*Centre for Nano Science and Engineering, Indian Institute of Science, Bangalore, 560012, India*

**S1: Device Fabrication: MoS$_2$ based nanoelectromechanical resonators**

The resonator is fabricated on a sapphire substrate, an insulating substrate, which helps in reducing parasitic capacitance, thus enabling direct capacitance measurements. The steps involved in the fabrication are schematically described in Figure S1 (a).

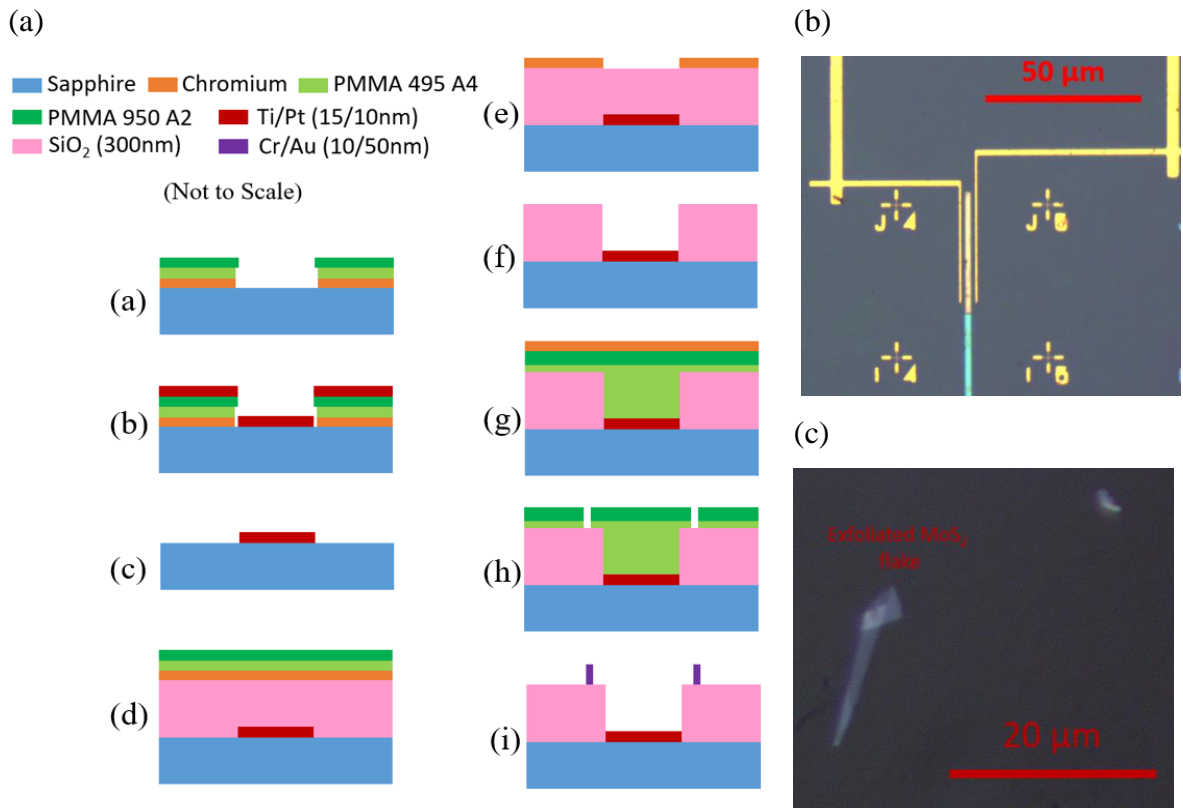

**Figure S1.** (a) Fabrication process flow for fabricating MoS$_2$ resonators. (b) Optical image (top view) of the fabricated device structure over which the MoS$_2$ flake is stamped to obtain the 2D beam resonator. (c) Exfoliated MoS$_2$ flake on PDMS stamp.

(a) Chromium deposition to reduce charging (during electron beam lithography (EBL) and patterning using EBL to define the local gate. (b) Cr etching from the developed region followed by Ti/Pt (15/10 nm) metal stack deposition as the gate electrode. (c) Lift-off of Ti/Pt metal followed by etching of remaining Cr to obtain a gate electrode over the sapphire substrate. (d) Deposition of $SiO_2$ using PECVD, conducting chromium layer, and spin coating of resist layers. (e) Patterning of gate region and gate pads followed by etching of Cr from the patterned area. Cr act as a mask for etching in the next step. (f) Reactive ion etching (RIE) to etch $SiO_2$ over the circular gate region to obtain the trench and over gate pads to enable bonding over the gate metal at a later stage, followed by etching of masking Cr layer. (g) Spin coating of resist layers and sputtering of conducting Cr layer for the next step of EBL. (h) Electron beam lithography for patterning of source-drain electrodes. (i) Deposition of Cr/Au (10/50nm) as source-drain electrodes followed by metal lift-off in acetone to obtain the final structure. Figure S1 (b) shows the optical image of the final structure obtained on which the $MoS_2$ flake is transferred. The transfer is done such that $MoS_2$ flake is suspended over the trench and makes contact to the source-drain electrodes. The transfer process makes use of the viscoelastic stamping process[1]. The $MoS_2$ flakes to be transferred is exfoliated on the PDMS stamp (Gel Pak) as shown in Figure S1 (c).

# Extended Data for Device (S2 – S4)

## S2: Dispersion Hysteresis with the direction of gate bias $(V_g^{dc})$ for MoS$_2$ beam resonator

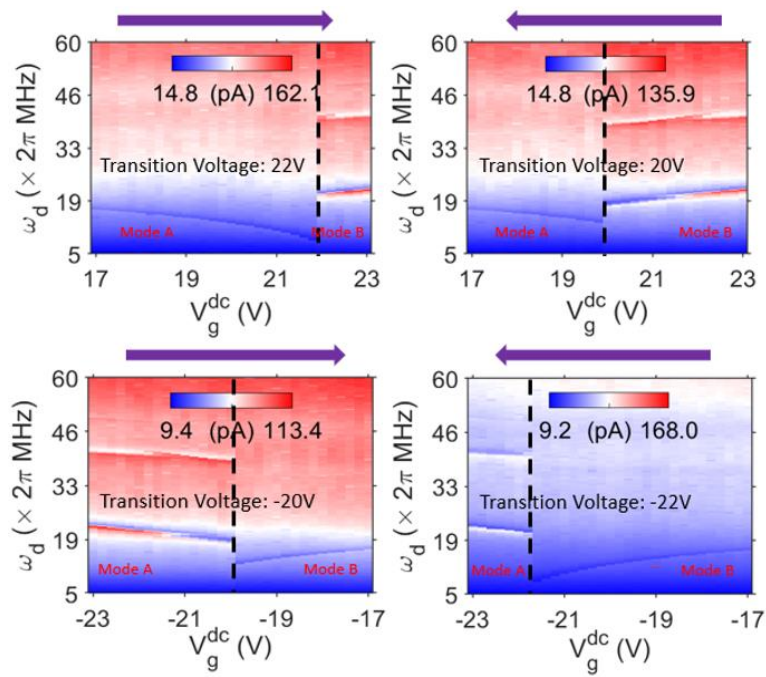

**Figure S2.** Dispersion plot and observation of hysteresis with gate bias for both positive and negative $V_g^{dc}$. The arrow above the plot shows the direction of the gate sweep. The measurement was carried out at a source-drain bias ($V_s^{ac}$) of 20mV and AC gate bias ($V_g^{ac}$) of 50mV.

In the positive direction of bias ($V_g^{dc}$) sudden transition from Mode A to B happens at 22V when starting from a lower gate bias and moving to a higher one. On reversing the sweep direction from higher gate bias to lower one (i.e., from 23V to 17 V), the transition from mode B to A happens at 20V. Similar behavior is observed in the negative side of the dispersion plot as well. This hysteretic phenomenon is related to poor clamping stability[3] and changes in clamping boundary conditions.

## S3: Phase response of Mode B and C (MoS₂ beam resonator)

Figure S3 (a) shows the phase response of Mode B and C measured at gate bias $V_g^{dc} = 21V$. Peak splitting is observed in the frequency response curve of mode B, indicating the intermodal exchange of energy. Figure S3 (b) and (c) shows the phase response Mode B and C measured at gate bias $V_g^{dc} = 23V$ and $25V$ respectively. At these gate voltages, no peak splitting is observed, indicating an inefficient exchange of energy among modes.

(a) $V_g^{dc} = 21V$

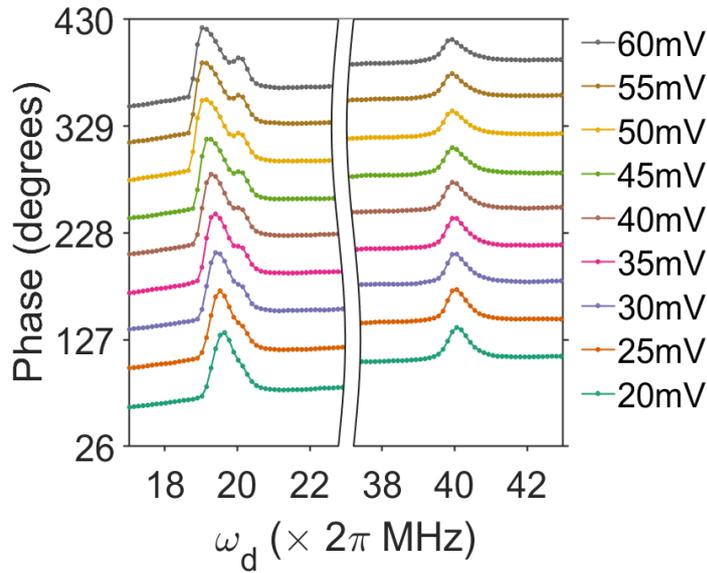

(c) $V_g^{dc} = 23V$

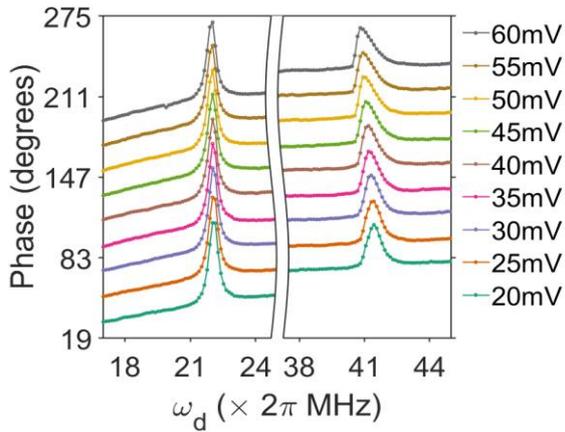

(c) $V_g^{dc} = 25V$

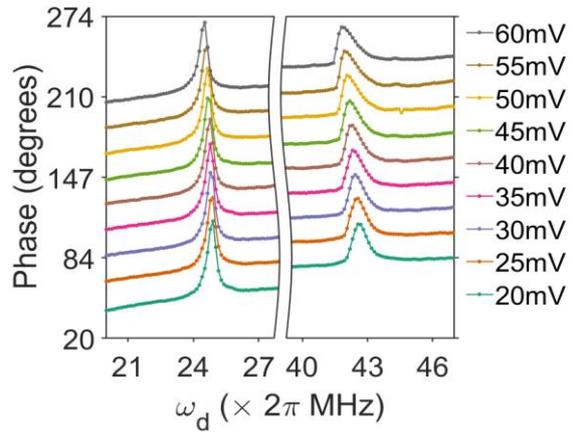

**Figure S3.** Phase response of Mode B and C measured at gate bias (a) $V_g^{dc} = 21V$ (b) $V_g^{dc} = 23V\ and$ (c) $V_g^{dc} = 25V$

## S4: Forward and Backward Frequency Sweep Response

(For varying gate voltages ($V_g^{dc}$) and ac drive ($V_g^{ac}$) amplitudes)

$V_g^{dc} = 21V$

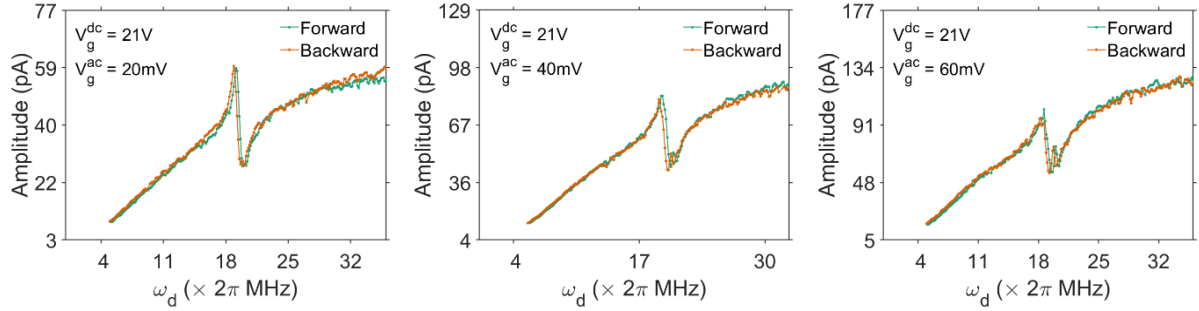

$V_g^{dc} = 23V$

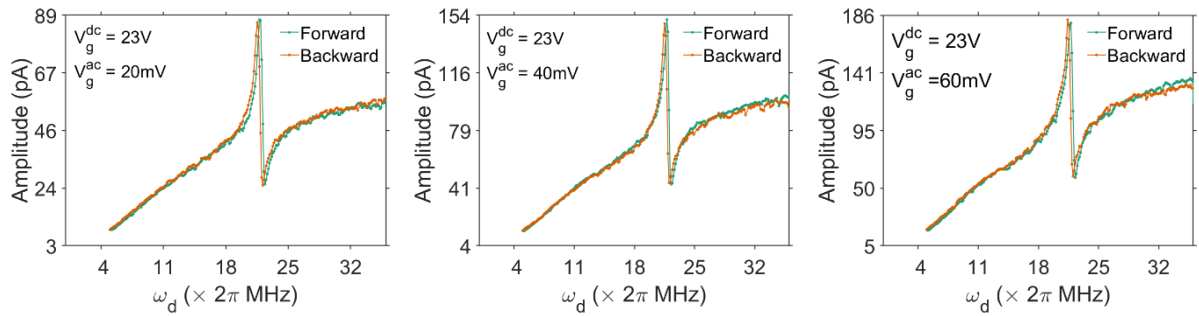

$V_g^{dc} = 25V$

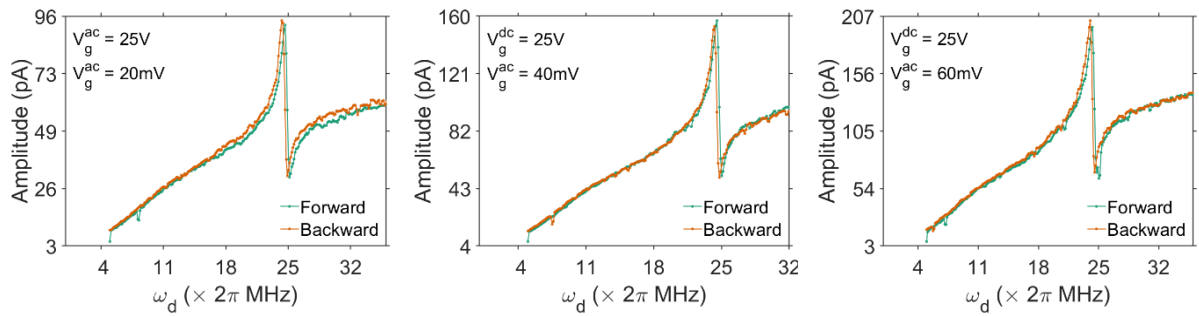

No apparent hysteresis is observed in the forward and backward frequency sweep (for the measurements mentioned in the main text). Hence, we conclude that the device has not yet entered the strongly nonlinear regime.

## S5: General derivation: Equation of motion for the quadratically coupled system using the Lagrangian approach

The Lagrangian approach is considered as the most general method of treating dynamical systems in terms of generalized coordinates.

The Lagrangian ($L$) is defined as $L = T - V$, and is a function of both the generalized coordinates and their velocities. Here $T$ is the total kinetic energy, and V is the total potential energy of the system. The equation of motion of any generalized system using the Lagrangian approach is defined using the following equation[4]:

$$\frac{d}{dt}\frac{\partial L}{\partial \dot{x}_j} - \frac{\partial L}{\partial x_j} - Q_j = 0,$$

Where, $x_j$ are the generalized coordinates and $Q_j$ is the generalized force acting on the system.

We model our system as two linear resonators coupled through the nonlinear quadratic coupling.

The interaction potential for the two systems is given as $V_{int} = x_1 x_2^2 + x_2 x_1^2$.

Total kinetic energy, $T = \frac{1}{2}m_1 \dot{x}_1^2 + \frac{1}{2}m_2 \dot{x}_2^2$

Total potential energy, $V = \frac{1}{2}m_1 \omega_1^2 x_1^2 + \frac{1}{2}m_1 \omega_1^2 x_1^2 + V_{int}$

The first two terms in the potential will have to be expanded further to account for nonlinearity in the system. In that case, the potential will not be harmonic.

$V = \frac{1}{2}m_1 \omega_1^2 x_1^2 + \frac{1}{2}m_1 \omega_1^2 x_1^2 + x_1 x_2^2 + x_2 x_1^2$, $Q_1 = F_1 Cos(\omega_d t)$, $Q_2 = F_2 Cos(\omega_d t)$

Thus, the equation of motion for our system using the above generalized equation comes out to be as follows:

$$\ddot{x}_1 + \omega_1^2 x_1 + 2\mu_1 \dot{x}_1 + \alpha_1 x_1^3 + \gamma_{22} x_2^2 + 2\gamma_{12} x_1 x_2 = \frac{F_1}{m_1} Cos(\omega_d t)$$

$$\ddot{x}_2 + \omega_2^2 x_2 + 2\mu_2 \dot{x}_2 + \alpha_2 x_2^3 + \gamma_{11} x_1^2 + 2\gamma_{21} x_1 x_2 = \frac{F_2}{m_2} Cos(\omega_d t)$$

We have introduced damping and effective cubic nonlinearity in the equation as well for the sake of completion and to lay down the basis for the most general model explaining modal coupling through 2:1 internal resonance.

Here, $x_1, x_2$ are displacements of the first and second mode respectively, $\mu_1, \mu_2 \approx \mu$ is the modal damping coefficient, $\alpha_1, \alpha_2$ is the effective nonlinearity of the two modes. This arises due to the additional tension produced in the membrane when it vibrates with a large amplitude and also due to the broken symmetry of the resonator caused due to the capacitive attraction of the gate. $\gamma_{11} \approx \gamma_{22}, \gamma_{12} \approx \gamma_{21}$ are the coupling coefficients, $\omega_1$ and $\omega_2$ are the resonant frequency of the two modes, respectively, $\omega_d$ is the driving force at which force $F_1$ and $F_2$ is applied to the first and second mode respectively.

## S6: Simulations using MATLAB for Dynamic Device Response (Runge Kutta Method: ODE 45)

Non-dimensional parameters used for simulating the dynamic response of modes using the coupled equation are mentioned along with the figures:

$\mu_1 \approx \mu_2 = 0.05$ is kept constant throughout all simulations.

## S6.1 Frequency Response (linear regime) in the absence of any coupling between the modes (on driving the first mode)

(a) First mode    (b) Second mode

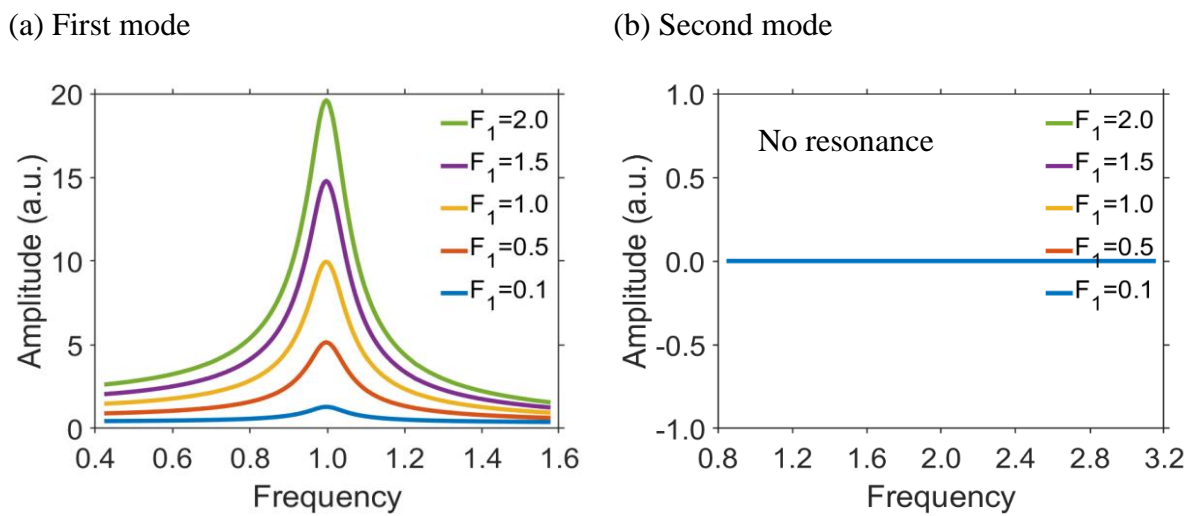

**Figure S6.1.** The simulated frequency response of (a) first mode and (b) second mode when the first mode is being driven. Since the two mode are uncoupled, no energy transfer from the first mode to the second takes place and is confirmed through the amplitude response of the second mode.

Simulation Parameters: $\alpha_1, \alpha_2 = 0, \gamma_{12} \approx \gamma_{11} = 0, \gamma_{21} \approx \gamma_{22} = 0, \omega_1 = 1.00, \omega_2 = 2.05$

## S6.2 Response of Mode 2 on Frequency detuning from 2: 1 Internal Resonance Condition

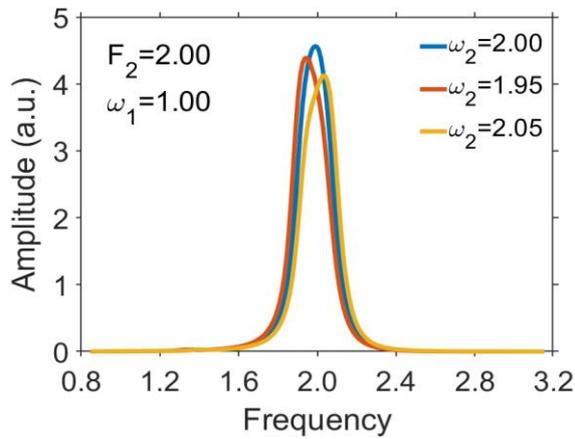

**Figure S6.2.** Effect (on mode 2) of detuning ratio of resonant frequencies such that $\frac{\omega_2}{\omega_1}\ ratio\ (= 2.0, 1.95, 2.05)$ is changed by varying frequency of $\omega_2$ keeping force ($F_1 = 2.00$) and other parameters constant.

Simulation Parameters: $\alpha_1, \alpha_2 = 0, \gamma_{12} \approx \gamma_{11} = 0.01,\ \gamma_{21} \approx \gamma_{22} = 0.03, F_1 = 2.00$

## S6.3 Response of Mode 1 on varying Coupling strength (more detuned frequency ratio as compared to Figure 5(b))

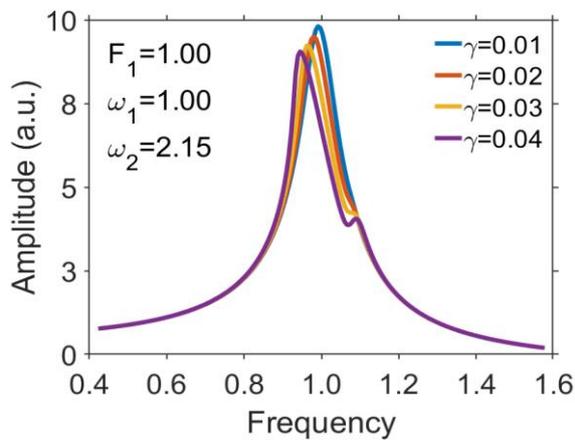

**Figure S6.3.** Effect of coupling strength on the frequency response of mode 1 ($\omega_1 = 1.00$) for more detuning from IR condition ($\omega_2 = 2.15$). Splitting in the peak is observed to increase with increasing coupling strength for the same forcing ($F_1 = 1.00$). But, in comparison to the less detuned case (manuscript Figure 5(b)) the peak splitting is much less for the same forcing and coupling strengths.

For simplicity we assume coupling as $(\gamma_{12} \approx \gamma_{11} \approx \gamma_{22} \approx \gamma_{21} = \gamma)$

# References

1. Castellanos-Gomez, A. *et al.* Deterministic transfer of two-dimensional materials by all-dry viscoelastic stamping. *2D Mater.* **1**, (2014).

2. Chen, C. Graphene NanoElectroMechanical Resonators and Oscillators. *Thesis Columbia University* (2013).

3. Aykol, M. *et al.* Clamping instability and van der Waals forces in carbon nanotube mechanical resonators. *Nano Lett.* **14**, 2426–2430 (2014).

4. Goldstein, H., Poole, C. & Safko, J. Classical mechanics. 3rd. (2002).

5. Xu, Y. *et al.* Radio frequency electrical transduction of graphene mechanical resonators. *Appl. Phys. Lett.* **97**, 95–98 (2010).

**************************************************************************